\documentclass[prb, reprint,twocolumn,showpacs,superscriptaddress,nobibnotes,floatfix]{revtex4-1}  
\usepackage{graphicx}  
\usepackage{dcolumn}  
\usepackage{bm}        
\usepackage{amssymb}   
\usepackage{amsmath}
\usepackage{tikz}

\mathchardef\mhyphen="2D

\graphicspath{{./}}

\makeatletter
\newcommand\textsubscript[1]{\@textsubscript{\selectfont#1}}
\def\@textsubscript#1{{\m@th\ensuremath{_{\mbox{\fontsize\sf@size\z@#1}}}}}
\newcommand\textbothscript[2]{%
  \@textbothscript{\selectfont#1}{\selectfont#2}}
\def\@textbothscript#1#2{%
  {\m@th\ensuremath{%
    ^{\mbox{\fontsize\sf@size\z@#1}}%
    _{\mbox{\fontsize\sf@size\z@#2}}}}}

\DeclareRobustCommand*{\citen}[1]{%
  \begingroup
    \romannumeral-`\x 
    \setcitestyle{numbers}%
    \cite{#1}%
  \endgroup   
}

\makeatother

\begin{document}

\title{Infrared nano-spectroscopy and imaging of collective superfluid excitations in conventional and high-temperature superconductors}

\author{H. T. Stinson}
\email{hstinson@physics.ucsd.edu}
\affiliation{Department of Physics, University of California, San Diego, La Jolla, California 92093, USA}
\author{J. S. Wu}
\affiliation{Department of Physics, University of California, San Diego, La Jolla, California 92093, USA}
\author{B. Y. Jiang}
\affiliation{Department of Physics, University of California, San Diego, La Jolla, California 92093, USA}
\author{Z. Fei}
\affiliation{Department of Physics, University of California, San Diego, La Jolla, California 92093, USA}
\author{A. S. Rodin}
\affiliation{Department of Physics, Boston University, 590 Commonwealth Avenue, Boston, Massachusetts 02215, USA}
\author{B. C. Chapler}
\affiliation{Department of Physics, University of California, San Diego, La Jolla, California 92093, USA}
\author{A.~S.~Mcleod}
\affiliation{Department of Physics, University of California, San Diego, La Jolla, California 92093, USA}
\author{A. Castro Neto}
\affiliation{Graphene Research Centre, National University of Singapore, 6 Science Drive 2, Singapore 117546}
\author{Y. S. Lee}
\affiliation{Department of Physics, Soongsil University, Seoul 156-743, South Korea}
\affiliation{Department of Physics, University of California, San Diego, La Jolla, California 92093, USA}
\author{M. M. Fogler}
\affiliation{Department of Physics, University of California, San Diego, La Jolla, California 92093, USA}
\author{D. N. Basov}
\affiliation{Department of Physics, University of California, San Diego, La Jolla, California 92093, USA}

\date{June 2014}

\begin{abstract}
We investigate near-field infrared spectroscopy and superfluid polariton imaging experiments on conventional and unconventional superconductors. Our modeling shows that near-field spectroscopy can measure the magnitude of the superconducting energy gap in Bardeen-Cooper-Schrieffer-like superconductors with nanoscale spatial resolution. We demonstrate how the same technique can measure the $c$-axis plasma frequency, and thus the $c$-axis superfluid density, of layered unconventional superconductors with a similar spatial resolution. Our modeling also shows that near-field techniques can image superfluid surface mode interference patterns near physical and electronic boundaries. We describe how these images can be used to extract the collective mode dispersion of anisotropic superconductors with sub-diffractional spatial resolution.
\end{abstract}

\maketitle

\section{Introduction}
\label{sec:intro}

Plasma excitations in superconductors have a rich and varied history. In the late 1950's, Anderson showed that conventional superconductors do not allow for bulk plasma excitations at energies below $2\Delta$, the magnitude of the superconducting gap.~\cite{Anderson1958, Anderson1963} However, collective oscillations of the superfluid at low frequencies in both conventional and high-temperature (high-$T_c$) superconductors can be excited in the form of surface plasmons.~\cite{Buisson1994,Dunmore1995a} Superfluid surface plasmons are  ``high-$q$", meaning that their in-plane momentum $q$ satisfies the inequality $q>\omega /c$, where $\omega$ is the frequency of the mode and $c$ is the speed of light in vacuum. This momentum mismatch makes such modes impossible to observe in typical optical experiments,~\cite{Dawson1994, Zayats2005} unless one resorts to nano-fabricated structures enabling high-$q$ coupling.~\cite{Dunmore1995a, Buisson1994}

High-$T_c$ cuprate superconductors, layered materials composed of CuO\textsubscript{2} planes, also exhibit a bulk low-frequency excitation of the superfluid along the $c$ axis.~\cite{Josephson1962,Tamasaku1992,Shibauchi1994,Basov1994,Tsui1996} This mode is a collective oscillation of the Josephson tunneling current normal to the CuO\textsubscript{2} planes. The Josephson plasma resonance occurs in the superconducting state at a frequency $\omega_c$ that varies with doping.~\cite{freqnote} For all dopings, $\omega_c$ is lower than $\omega_{ab}$, the plasma frequency in the $ab$ plane.~\cite{Homes2004} This anisotropy of plasma frequencies allows for two families of superfluid surface modes that couple to the Josephson plasmon.~\cite{Doria1997, Guo2012} Measurements of the surface mode dispersion in high-$T_c$ superconductors can yield complete information on both the in-plane and the $c$-axis dielectric functions, $\epsilon_{ab}(\omega)$ and $\epsilon_c(\omega)$, respectively.  

In this work, we propose an experimental approach utilizing scattering-type scanning near-field optical microscopy \mbox{(s-SNOM)}~\cite{Pohl1984, Denk1991, Cvitkovic2007, Amarie2009,Novotny2006, Atkin2012} to directly probe the local variation of the superfluid response at the nanoscale and map the spectrum of collective superfluid surface modes. The rest of this paper is separated into four parts. In Sec.~\ref{sec:math}, we summarize and improve upon previously derived results on the collective mode spectrum of anisotropic superconductors, such as those in the cuprate family. In Sec.~\ref{sec:spi}, we describe how s-SNOM can map the dispersion of collective superfluid excitations in superconducting thin films or exfoliated crystals through real-space imaging measurements.~\cite{Chen2012f,Fei2012a, Fei2013} We refer to this technique as scanning plasmon interferometry (SPI). Our modeling predicts that SPI can map the dispersion of superfluid modes for the prototypical high-$T_c$ cuprate compounds  La\textsubscript{2-x}Sr\textsubscript{x}CuO\textsubscript{4} (LSCO) and  YBa\textsubscript{2}Cu\textsubscript{3}O\textsubscript{x} (YBCO). These dispersion maps can then be used to extract the anisotropic optical constants of the superconductor and their spatial variation at length scales much shorter than the wavelength of light at the probing frequency. In Sec.~\ref{sec:spec}, we explain how s-SNOM spectroscopy can compliment dispersion measurements. We establish that spectroscopic s-SNOM can extract the magnitude of the Jospehson plasma resonance frequency $\omega_c$, and thus the $c$-axis superfluid density, of anisotropic superconductors with nanoscale spatial resolution. Moreover, we show how s-SNOM spectroscopy can measure the superconducting energy gap in thin films of conventional superconductors with identical spatial resolution. We describe how the gap feature in s-SNOM spectra is related to the surface plasmon mode in superconducting thin films described by Bardeen-Cooper-Schrieffer (BCS) theory.. In Sec.~\ref{sec:conc}, we summarize our results, which show that s-SNOM is a powerful potential tool for probing both the superfluid density and the superconducting energy gap at ultrasmall length scales.

\section{Collective Modes in Anisotropic Superconductors}
\label{sec:math}

\subsection{Overview}

\begin{figure}[b]
\includegraphics{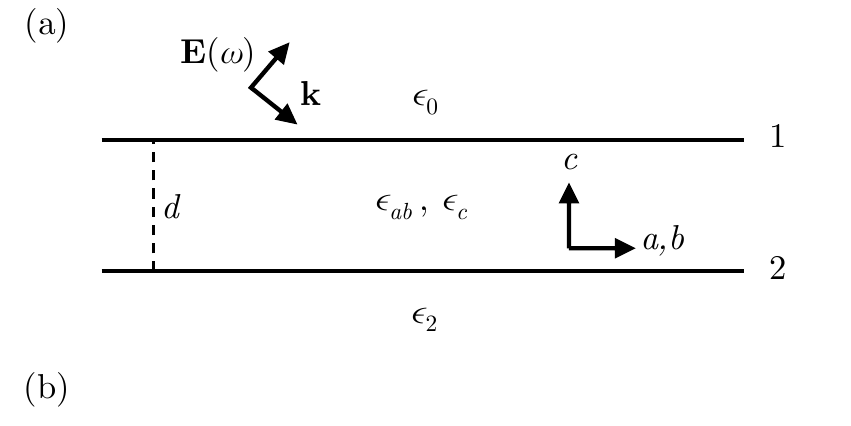}
\includegraphics{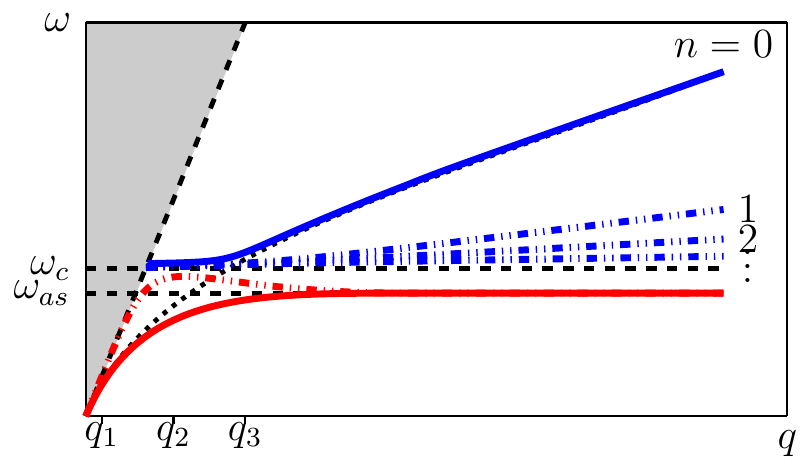}
\caption{(a) Schematic of the uniaxial thin film geometry under consideration. The dielectric constants above and below the thin film are $\epsilon_0$ and $\epsilon_2$, respectively. (b)~Schematic of the collective mode dispersion for an anisotropic superconducting thin film. All collective modes lie outside the light cone (shaded, $q<\sqrt{\epsilon_j}\omega /c$). The Josephson plasma frequency $\omega_c$ is shown by the upper dashed horizontal black line. The symmetric [lower red solid line] and antisymmetric [lower red dot-dashed line] SPP modes lie below $\omega_c$, and the principal [upper blue solid line; $n=0$] and higher-order [upper blue dot-dashed lines; $n=1$, $2,\ldots$] HWMs lie above it. Solid lines are the modes we consider for imaging; dot-dashed lines are modes that are likely unobservable. The dotted black line denotes continuation of the symmetric SPP mode into the principal HWM above $\omega_c$. $q_1$, $q_2$, and $q_3$ denote the boundaries of different dispersive behaviors for the three modes, and the lower dashed horizontal black line at $\omega_{as}$ denotes the asymptotic SPP frequency at large $q$.
 \label{fig:colmod}}
\end{figure}

The collective mode spectrum of the cuprates has been derived previously.~\cite{Doria1997,Artemenko1995c,Slipchenko2011} Here, we provide a brief overview, starting with uniaxial materials in general.~\cite{Azzam1977} Collective electromagnetic modes in a material correspond to poles of the reflectivity coefficient $r_P$, which depends on both material constants and geometry.~\cite{Zhang2012} For uniaxial materials, the dielectric function is the matrix
\begin{equation}
\tilde{\epsilon}=
\begin{pmatrix}
\epsilon_{ab}&0&0\\
0&\epsilon_{ab}&0\\
0&0&\epsilon_c
\end{pmatrix}\,.
\end{equation}
In what follows, we consider $P$-polarized electromagnetic radiation incident on a thin film of uniaxial crystal with thickness $d$, in-plane dielectric function $\epsilon_{ab}$, and $c$-axis dielectric function $\epsilon_c$, with the $c$-axis parallel to the plane normal. Figure~\ref{fig:colmod}(a) shows a schematic of the system under consideration. The reflectivity coefficient $r_P$ as a function of the incident frequency $\omega$ and in-plane wavevector $q$ is given by 
\begin{equation}
\label{eq:rp}
r_P\left(\omega,q\right)=\frac{r_1+r_2e^{2ik_1d}}{1+r_1r_2e^{2ik_1d}}, 
\end{equation}
in which $r_1$ and $r_2$ stand for
\begin{subequations}
\begin{align}
\label{eq:rpgroup}
r_1 &=\frac{\epsilon_{ab}k_0-\epsilon_0k_1}{\epsilon_{ab}k_0+\epsilon_0k_1}\\
r_2 &=\frac{\epsilon_2k_1-\epsilon_{ab}k_2}{\epsilon_2k_1+\epsilon_{ab}k_2},
\end{align}
\end{subequations}
the reflection coefficients at the first and second interfaces. The parameter $\epsilon_0$ ($\epsilon_2$) is the dielectric constant of the material above (below) the thin film, 
\begin{equation}
k_j=\sqrt{\epsilon_j\left(\frac{\omega}{c}\right)^2-q^2},\quad j=0,2
\end{equation}
 is the component of the wave vector in the medium above $(j=0)$ or below $(j=2)$ the thin film perpendicular to the interface, and
\begin{equation}
k_1=\sqrt{\epsilon_{ab}\left(\frac{\omega}{c}\right)^2-\frac{\epsilon_{ab}}{\epsilon_c}q^2}
\end{equation}
is the component of the wave vector in the thin film perpendicular to the interface. Collective modes correspond to poles of $r_P$. A straightforward derivation~\cite{Azzam1977} shows that the poles of $r_P$ are the solutions to
\begin{equation}
\text{tanh}\left(ik_1d\right)=\frac{\epsilon_{ab}\left(\epsilon_2k_0k_1+\epsilon_0k_1k_2\right)}{\epsilon_0\epsilon_2k_1^2+\epsilon_{ab}^2k_0k_2}\,.
\label{eq:drfull}
\end{equation}
We look for solutions that decay exponentially into either dielectric, which requires that the imaginary parts of $k_{0,2}$ are positive. This property corresponds to modes confined to the interface of the thin film and the dielectric.

There are in general three frequency regions where solutions to Eq.~\eqref{eq:drfull} exist, characterized by the relative signs of $\mathrm{Re}\,\epsilon_{ab}$ and $\mathrm{Re}\,\epsilon_c$. When both $\mathrm{Re}\,\epsilon_{ab}$ and $\mathrm{Re}\,\epsilon_c$ are negative, $k_1$ is purely imaginary. The corresponding solutions of Eq.~\eqref{eq:drfull} describe surface plasmon polaritons (SPP) modes, which decay exponentially into both the surrounding dielectric and the thin film. The higher frequency SPP branch in cuprate systems is likely unobservable due to the relatively large anisotropy of most high-$T_c$ superconductors, so in our modeling we focus on the lower frequency SPP branch. In the case where $\epsilon_0=\epsilon_2$, the higher and lower SPP modes are referred to as antisymmetric and symmetric, respectively.~\cite{Doria1997} Both SPP modes approach the asymptotic frequency $\omega_{as}$ for large $q$. At frequencies $\omega>\omega_c$, $\mathrm{Re}\,\epsilon_{ab}$ is negative while $\mathrm{Re}\,\epsilon_c$ is positive, implying $k_1$ has a finite real part. The modes in this frequency range are hyperbolic waveguide modes (HWMs), which propagate in the thin film, but decay exponentially in the surrounding dielectrics.~\cite{Doria1997,Slipchenko2011} There are infinitely many HWMs, denoted by the index $n$. In the context of plasmon imaging experiments, we focus on the $n=0$ or principal HWM, which can be thought of as the continuation of the symmetric SPP to frequencies above $\omega_c$, as shown in Fig.~\ref{fig:colmod}(b). We do not consider the case when $\omega>\omega_{ab}$ and $\mathrm{Re}\,\epsilon_{ab},\mathrm{Re}\,\epsilon_c>0$, because this condition is typically realized at frequencies much higher than the superconducting energy gap $2\Delta$. A schematic of the collective mode dispersion near $\omega_c$ is shown in Fig.~\ref{fig:colmod}(b). In previous works,~\cite{Buisson1994,Dunmore1995a} both the SPP and the HWM are collectively referred to as two-dimensional (2D) plasmons. These two families of surface modes are presumably observable only in films or exfoliated crystals with thicknesses less than roughly $100$ nm. All of these plasmonic modes are overdamped in the normal state.

\subsection{Collective mode dispersion relations}

As the first approximation, we use the London model to describe the in- and out-of-plane components of the dielectric tensor of a layered superconductor,
\begin{align}
\label{eq:london}
\epsilon_c &=\epsilon_{\infty}\left(1-\frac{\omega_c^2}{\omega^2}\right),\\
\epsilon_{ab} &= \epsilon_{\infty}\left(1-\frac{\gamma^2\omega_c^2}{\omega^2}\right),
\end{align}
where $\gamma>1$ is a dimensionless parameter describing the anisotropy of the material. The London model omits both the increase in the real part of the optical conductivity above $2\Delta$ due to the breaking of Cooper pairs and any contribution to Re~$\sigma$ at frequencies below $2\Delta$ due to a residual normal-state fluid. However, the London model is sufficient to describe the general character of collective modes in the layered cuprates, as these omissions will primarily contribute to damping of the modes.

\subsubsection{Surface plasmon polariton modes in layered superconductors}
SPPs are confined to the two interfaces of the thin film and the dielectric. If the decay length $2\pi /k_1$ of the SPP away from the interface is greater than the film thickness $d$, the two surface modes can couple. This coupling leads to a splitting of the SPP mode into two branches,as described above.~\cite{Doria1997} In what follows, we consider a symmetric dielectric environment where $\epsilon_0=\epsilon_2$. In this case we can refer to the upper and lower frequency branches as the antisymmetric and symmetric modes. Their dispersions are both photon-like for small $q$. With increasing $q$, the coupling causes a level repulsion between the two modes. For extremely large $q$, the two surfaces decouple and both of the modes have the same frequency.  In the sequence of increasing $q$, the SPP regions are called $\it{optical, coupled,}$ and $\it{asymptotic}$.~\cite{Doria1997} We find the asymptotic frequency for large $q$ to be
\begin{align}
\omega_{as}=\left[\frac{2\gamma^2}{1+\gamma^2+\sqrt{(\gamma^2-1)^2+4\gamma^2\frac{\epsilon_{0}^2}{\epsilon_{\infty}^2}}}\right]^{1/2}\omega_{c}.
\label{eq:was}
\end{align}

A somewhat different formula for $\omega_{as}$ was given in Ref.~\citen{Doria1997}.  We believe our expression to be the correct one. For $\gamma=1$, 
\begin{equation}
\omega_{as}=\dfrac{\omega_c}{\sqrt{\left(1+\dfrac{\epsilon_0}{\epsilon_{\infty}}\right)}}\,,
\end{equation}
the surface plasmon frequency for a single interface.~\cite{Raether1965,Maier2007} As $\gamma$ increases, $\omega_{as}$ approaches the Josephson plasma frequency.

The three characteristic $q$'s that label the boundaries of the optical, coupled, and asymptotic regions of the surface plasmon mode dispersion are (for $\gamma\gg1$)
\begin{align}
q_{1}&=\gamma^2d\epsilon_{\infty}\dfrac{\omega_c^2}{c^2},\\
q_{2}&=\dfrac{\epsilon_{\infty}\omega_c}{c\sqrt{\epsilon_{0}}},\\
q_{3}&=\dfrac{\epsilon_{0}}{\epsilon_{\infty}}\dfrac{1}{\gamma^2d}, \label{eq:q3}
\end{align}   
For the symmetric mode, the dispersions in the optical, coupled, and asymptotic regions read
\begin{align}
\omega\approx\left\{
\begin{array}{ll}
\dfrac{cq}{\sqrt{\epsilon_{0}}}&,~~q\ll q_{1}\\ \\
\gamma\omega_{c}\sqrt{\dfrac{1}{1+\dfrac{\epsilon_{0}}{\epsilon_{\infty}}\dfrac{2}{qd}}}&,~~q_{1}\ll q\ll q_{3}\\ \\
\omega_{as}&, ~~q_{3}\ll q 
\end{array}
\right.\label{eq:sym}
\end{align}
For the antisymmetric mode, the dispersions read
\begin{align}
\omega\approx\left\{
\begin{array}{ll}
\dfrac{cq}{\sqrt{\epsilon_{0}}}&,~~q\ll q_{2}\\ \\
\omega_{c}\sqrt{\dfrac{1}{1+\dfrac{\epsilon_{0}}{\epsilon_{\infty}}\dfrac{qd}{2}}}&,~~q_{2}\ll q\ll q_{3}\\ \\
\omega_{as}&, ~~q_{3}\ll q
\end{array}
\right.\label{eq:antisym}
\end{align}
First, we note that the antisymmetric surface mode reaches a maximum frequency of $\omega_c$ at $q_2$, and then asymptotically approaches $\omega_{as}$. However, for materials with large $\gamma$, such as most cuprates, $\omega_{as}\approx\omega_c$. Thus, the antisymmetric SPP's coupled region (where the mode dispersion curve is clearly separated from both the light cone and the Josephson plasma) is small and, most likely, unobservable. Second, the assumed inequality $q_1<q_2$ is satisfied only if  $\gamma^2 d<c \epsilon_{\infty}/\omega_c\epsilon_0$. In the cuprates,~\cite{Homes2004} $\omega_c$ is typically of the order of 10\textendash100 cm\textsuperscript{-1} and $\gamma$ is of the order of 10\textendash100. This means that $q_1<q_2$ only for films with thicknesses of the order of nanometers.  In thicker films, the intermediate ``coupled'' regime is absent.

\subsubsection{Hyperbolic waveguide modes}
\begin{figure*}
\includegraphics{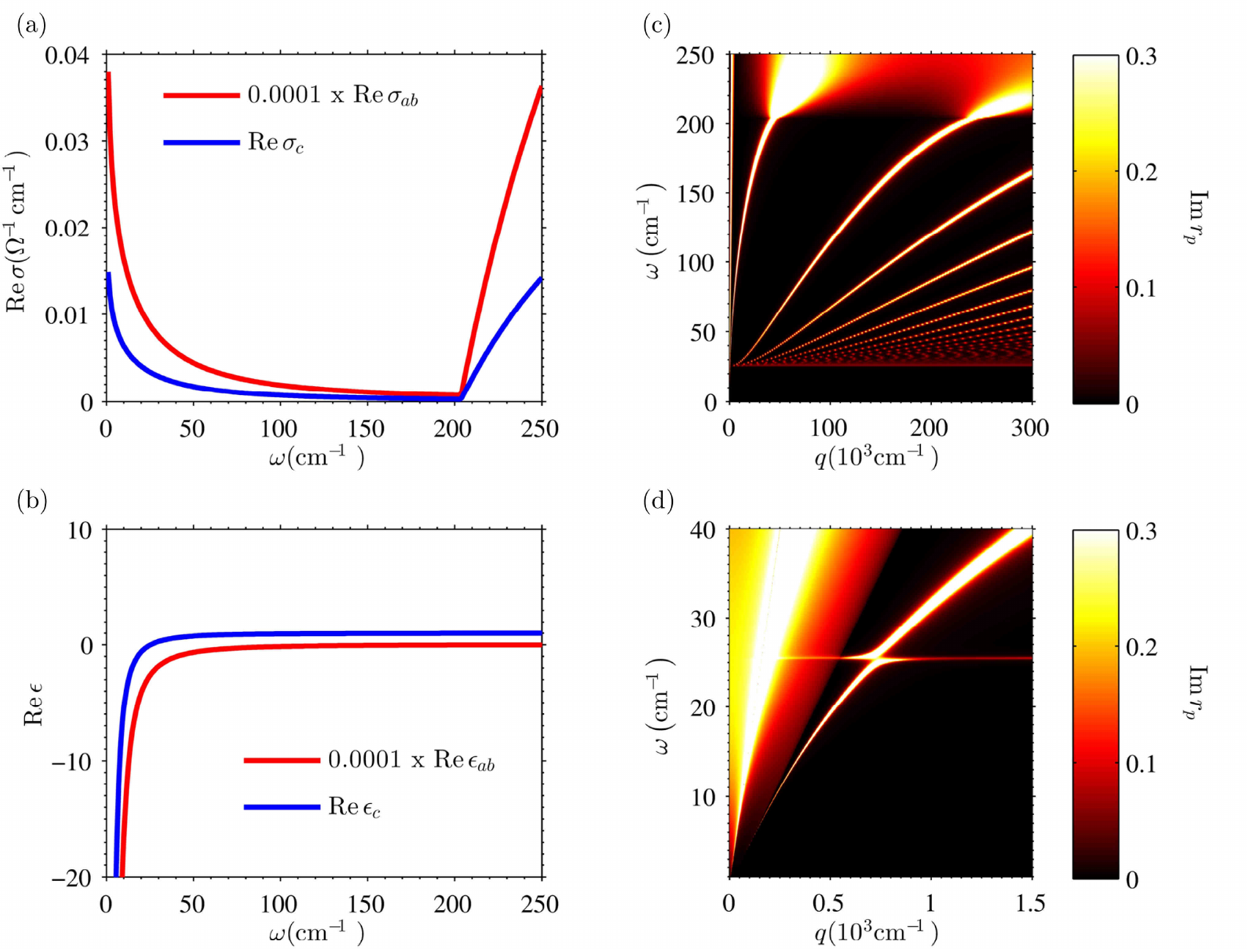}
\caption{(a) The real part of the optical conductivity and (b) the real part of the dielectric function for a BCS-like anisotropic superconductor model. The unscreened plasma frequencies are $\omega_{ab}=8000\textrm{ cm}^{-1}$ and $\omega_c=50\textrm{ cm}^{-1}$, and both directions have $2\Delta=200\textrm{ cm}^{-1}$, scattering frequency $\beta=800\textrm{ cm}^{-1}$, and $T/T_c=0.37$. (c) The imaginary part of the reflection coefficient for a 10-nm thin film on a Si substrate as a function of $\omega$ and $q$, calculated from Eq. \eqref{eq:rp} using the BCS optical constants in (a) and (b). (d) Expanded view of (c) at low $\omega$ and $q$ shows continuation of the symmetric SPP mode into the principal HWM at $\omega=\omega_{as}$.
\label{fig:bcsrp}}
\end{figure*}
\begin{figure*}
\includegraphics{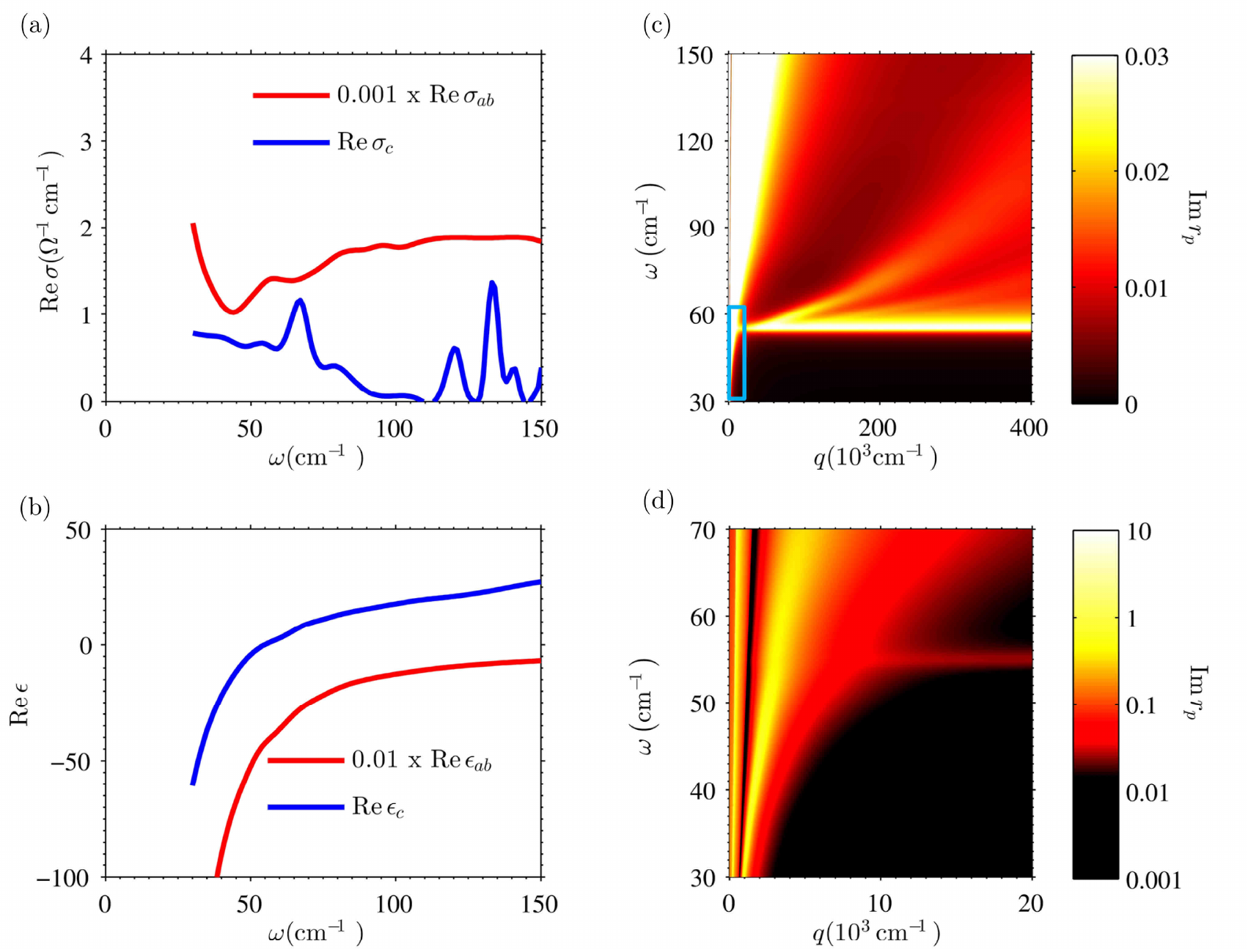}
\caption{(a) The real part of the optical conductivity and (b) the real part of the dielectric function for both the $ab$ plane and the $c$ axis of LSCO ($x=0.15$, $T\ll T_c$), taken from reflectivity measurements with low-frequency $c$-axis phonons subtracted.~\cite{Tajima2005, Uchida1996} (c) The imaginary part of the reflection coefficient for a 10-nm LSCO film ($x=0.15$, $T\ll T_c$) on a Si substrate as a function of $\omega$ and $q$, calculated from Eq.~\eqref{eq:rp} using the LSCO optical constants in (a) and (b). The bright horizontal line at $\omega_c\approx 55 \textrm{ cm}^{-1}$ is due to the Josephson plasma resonance. The upward-bending lines above $\omega_c$ at $q\approx 1.0 \times 10^5\,\mathrm{cm}^{-1}$ are the higher-order HWMs. The rapidly increasing width of these lines is due to damping from the high residual conductivity at these frequencies. At $T>T_c$, all of these modes are overdamped and no longer visible. (d) Expanded view of the small blue rectangular box in (c), showing the symmetric SPP mode and the principal HWM. The crossover from the SPP into the principal HWM is blurred by damping in the $ab$ plane. The logarithmic color scale emphasizes the much stronger absorption of these modes than the higher-order HWMs in (c). The horizontal red line is the Josephson plasma resonance.
\label{fig:lscorp}}
\end{figure*}
\begin{figure}
\includegraphics{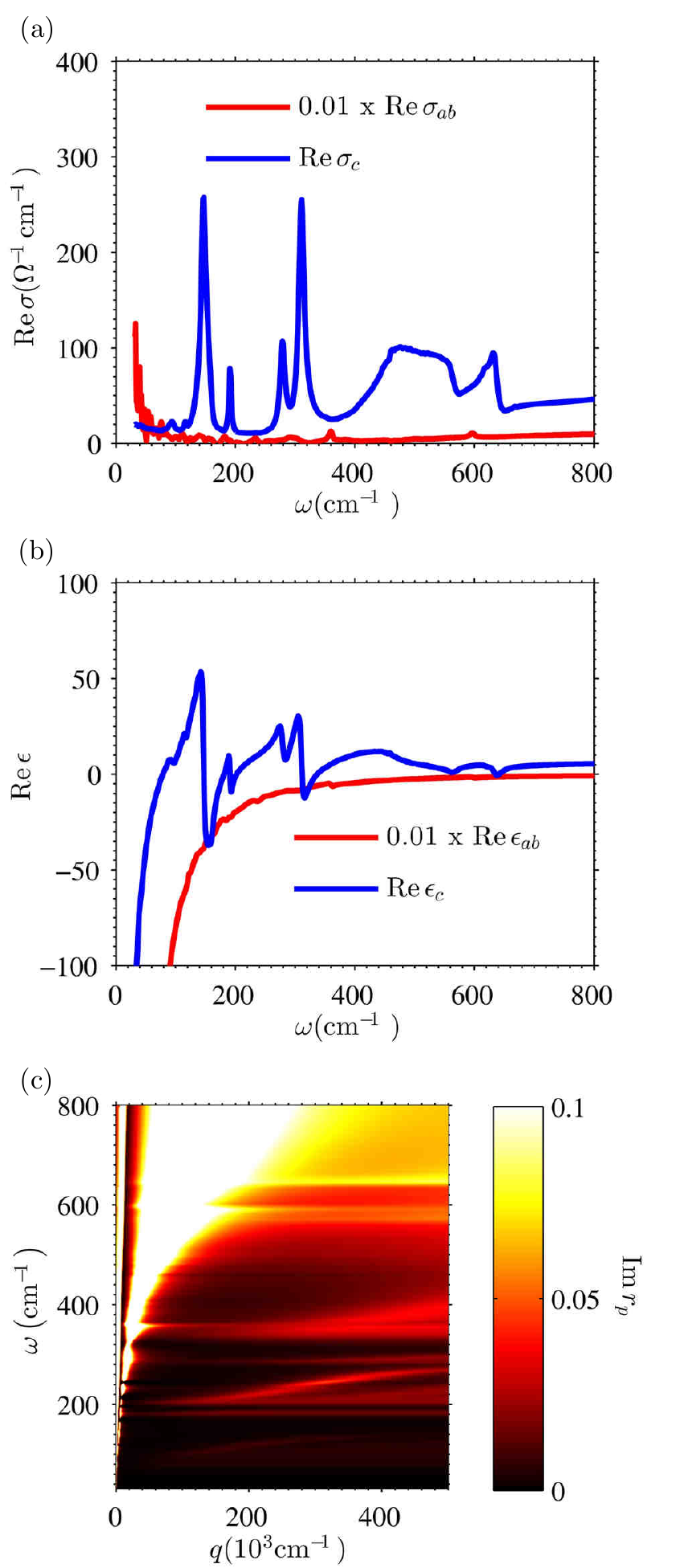}
\caption{(a) The real part of the optical conductivity and (b) the real part of the dielectric function for both the $ab$ plane and the $c$ axis of YBCO ($x=6.75$, $T\ll T_c$), taken from Refs.~[\citen{Lee2005}] and~[\citen{Homes1995}]. (c) The imaginary part of the reflection coefficient for a 10-nm YBCO film on a Si substrate as a function of $\omega$ and $q$, calculated using the YBCO optical constants in (a) and (b). The collective modes are hybridized with the phonon resonances below the gap; the YBCO $r_P$ is otherwise very similar to that of the phonon-subtracted LSCO [Fig.~\ref{fig:lscorp}(c)].
\label{fig:ybcorp}}
\end{figure}

At $q\gg\sqrt{\epsilon_j}\,\omega/c$, Eq.~\eqref{eq:drfull} reduces to
\begin{eqnarray}
q=&&\frac{i}{d}\frac{\sqrt{\epsilon_{c}}}{\sqrt{\epsilon_{ab}}}\Biggl[n\pi+\arctan
\left(\frac{\epsilon_{0}}{-i\sqrt{\epsilon_{ab}}\sqrt{\epsilon_{c}}}\right)\nonumber\\
&&+\arctan\left(\frac{\epsilon_{2}}{-i\sqrt{\epsilon_{ab}}\sqrt{\epsilon_{c}}}\right)
\Biggr],
\end{eqnarray}
which in the case of a symmetric environment $(\epsilon_0=\epsilon_2)$, becomes
\begin{equation}
q=\frac{2i}{d}\frac{\sqrt{\epsilon_{c}}}{\sqrt{\epsilon_{ab}}}\left[\frac{n\pi}{2}+\arctan
\left(\frac{\epsilon_{0}}{-i\sqrt{\epsilon_{ab}}\sqrt{\epsilon_{c}}}\right)\right],
\label{eq:wgm}
\end{equation}
with $n=0,~1,~2,\ldots$. For $n\geq1$ and $\omega>\omega_c$, the arctangent term in Eq.~\eqref{eq:wgm} is negligible. The dispersion of HWMs for $n\geq1$ then simplifies to  
\begin{equation}
\omega\approx\omega_{c}\sqrt{\frac{\gamma^2\left(\frac{qd}{n\pi}\right)^2+1}{\left(\frac{qd}{n\pi}\right)^2+1}},
\end{equation}
which agrees with the dispersion found in Ref.~\citen{Artemenko1995c}. However, the formula for the dispersion of the $n=0$ or principal HWM given in Ref.~\citen{Artemenko1995c} is not correct for $q\gg q_{3}$. One can show that the dispersion of the principal HWM has a sharp inflection at $q_{3}$ due to the arctangent term in Eq.~\eqref{eq:wgm}:
\begin{equation}
\omega\approx\left\{\begin{array}{cl}\omega_{c}\sqrt{\dfrac{\gamma^2\left(\dfrac{qd}{\pi}\right)^2+1}{\left(\dfrac{qd}{\pi}\right)^2+1}}~&,~~q\ll q_{3}\\ \\
\gamma\omega_{c}\sqrt{\dfrac{1}{\dfrac{\epsilon_{0}}{\epsilon_{\infty}}\dfrac{2}{qd}+1}}~&,~~q\gg q_{3}
\end{array}\right..\label{eq:n=0}
\end{equation} 

The second line of Eq.~\eqref{eq:n=0} is the same as the dispersion relation in the coupled region of the symmetric SPP mode. Thus, the principal HWM can be understood as the continuation of the symmetric surface mode to higher frequencies $\omega>\omega_{c}$. This dispersion has an approximately $\sqrt{q}$-dependence for $q<1/d$, typical of 2D plasmons. This form of $q$-dependence originates from the in-plane motion of the electrons, as evidenced by the fact that in the limit where $qd\ll1$ and we assume that $|\epsilon_{ab}|\gg|\epsilon_0|,|\epsilon_2|$ (valid for $\omega\ll\omega_{ab}$), Eq.~\eqref{eq:drfull} reduces to
\begin{equation}
q\cong\frac{i\kappa\omega}{2\pi d\sigma_{ab}}\,,
\label{eq:symdisp}
\end{equation}
where $\kappa=\dfrac{\epsilon_0+\epsilon_2}{2}$ is the average dielectric function of the surrounding medium. Here we have substituted the optical conductivity $\sigma_{ab}$ for the dielectric permittivity $\epsilon_{ab}$ using the relation 
\begin{equation}
\epsilon_{ab}=1+\frac{4\pi i\sigma_{ab}}{\omega}\,.
\end{equation}
Thus, in the high-$q$ limit the symmetric SPP and the principal HWM dispersion no longer depend on the $c$-axis optical constants.

\subsection{Reflection coefficients of layered cuprates and conductivity models}

The London model in Eq.~\eqref{eq:london} fails to capture some essential features of measured cuprate optical constants, namely, the residual normal-fluid conductivity at energies below the gap and the sharp increase in dissipation at energies above the gap.~\cite{Basov2005} To better account for finite dissipation in real materials, we use optical constants calculated from a BCS model with finite scattering.~\cite{Zimmermann1991} To capture the anisotropy of high-$T_{c}$ superconductors, we assume different values of the screened plasma frequency in and out of the plane. Figures~\ref{fig:bcsrp}(a) and~\ref{fig:bcsrp}(b) show the real part of both the $ab$-plane and the $c$-axis optical constants calculated for the BCS model. Both the in- and the out-of-plane optical constants were calculated with a gap magnitude $2\Delta=200\,\mathrm{cm}^{-1}$. Figure \ref{fig:bcsrp}(c) shows the imaginary part of $r_P$, which is maximized along the dispersion curves of the collective modes.~\cite{Zhang2012} The HWMs are clearly visible for $\omega>\omega_c$. As the frequency increases above $2\Delta$, the collective mode dispersions rapidly become incoherent. The sharp collective mode resonance transitions into a broad dissipative background at all wavevectors. This mode broadening at the gap permits s-SNOM measurements of the energy gap magnitude, as discsussed in Sec.~\ref{sec:spec}. Finally, Fig.~\ref{fig:bcsrp}(d) shows the crossing of the symmetric SPP mode into the principal HWM at $\omega=\omega_{as}$.

Although the anisotropic BCS model can help us understand the behavior of collective modes in cuprate thin films, it is not fully realistic. Therefore, we also calculate $r_P$ using measured optical constants for LSCO and YBCO. Figures~\ref{fig:lscorp} and~\ref{fig:ybcorp} show the optical constants and imaginary part of $r_p$ for LSCO and YBCO, respectively. The LSCO optical constants for $x=0.15$ at $T=5$K are taken from Ref.~\citen{Tajima2005} for the $ab$-plane and Ref.~\citen{Uchida1996} for the $c$-axis, with $c$-axis phonons subracted. The YBCO data for $x=6.75$ at $T=10$K are from Ref.~\citen{Lee2005} for the $ab$-plane, and the $c$-axis data were measured by the authors as described in Ref.~\citen{Homes1995}; phonons are not subtracted from the YBCO spectra.

In Figs.~\ref{fig:lscorp}(c) and~\ref{fig:lscorp}(d), the bright horizontal line near \mbox{$\omega=55$~cm\textsuperscript{-1}} is the asymptotic SPP mode at $\omega_{as}$, which, in the limit of large anisotropy~\cite{asympnote}, is approximately equal to the Josephson plasma frequency $\omega_c$. The significance of the Josephson plasma frequency is that it is directly related to the $c$-axis superfluid density $\rho_{c,s}$.~\cite{Tamasaku1992,Dordevic2003} Typically, spectroscopic observables of the Josephson plasma resonance fall in the far-infrared or terahertz (THz) frequency range.~\cite{Homes2004} The range of Im $r_P$ represented by the logarithmic color scale is much higher in Fig. \ref{fig:lscorp}(d), masking the horizontal line near 55 cm\textsuperscript{-1}. This indicates that the symmetric SPP and the principal HWM have a higher oscillator strength than the higher-order HWMs, meaning that the former will be much easier to excite. The high residual conductivity of the cuprates in the superconducting state broadens both these modes, blurring the crossover from SPP to HWM at $\omega_c$.

In both Fig.~\ref{fig:lscorp} and~\ref{fig:ybcorp}, the collective modes are much more damped than in the simple BCS model, a result of the high residual conductivity below $2\Delta$ in the cuprates. In Fig.~\ref{fig:lscorp}(c), the first few HWMs are visible at frequencies slightly above $\omega_c$, but are quickly damped as $q$ increases. The collective mode spectrum is clearly dominated by the  2D plasmon-like modes.  Figure~\ref{fig:lscorp}(d) shows the $\sqrt{q}$-like modes near $\omega_c$, but the large damping in the $ab$ plane smears out the crossing of the SPP mode into the principal HWM. We see essentially the same behavior in the YBCO data [Fig. \ref{fig:ybcorp}(c)]; the only difference being that the electromagnetic collective modes hybridize with the phonons, complicating the simple model sketched in Fig.~\ref{fig:colmod}(b). We conclude that in real materials, excitation of a coherent HWM for $n\geq1$ is challenging due to the high residual conductivity in the $c$ axis above the Josephson plasma frequency. However, the symmetric SPP mode and the principal HWM could be observable using high-$q$ probes such as s-SNOM.

\section{Surface Plasmon Interferometry of Layered Superconductors}
\label{sec:spi}

Both the symmetric SPP and the principal HWM have much higher momenta than photons at the same frequency in vacuum. In previous experiments on YBCO thin films, light was coupled to superfluid SPP modes via a patterned grating.~\cite{Dunmore1995a} Any subwavelength scatterer can provide the necessary momenta to couple to high-$q$ modes, including the apex of an s-SNOM tip. It has recently been shown that the s-SNOM apparatus can both excite and detect SPPs in graphene~\cite{Fei2012a,Chen2012f,Fei2013} and boron nitride.~\cite{Dai2013} This virtue of s-SNOM can also be exploited to measure the dispersion of the collective superfluid modes in cuprate thin films. 

\begin{figure}
\includegraphics{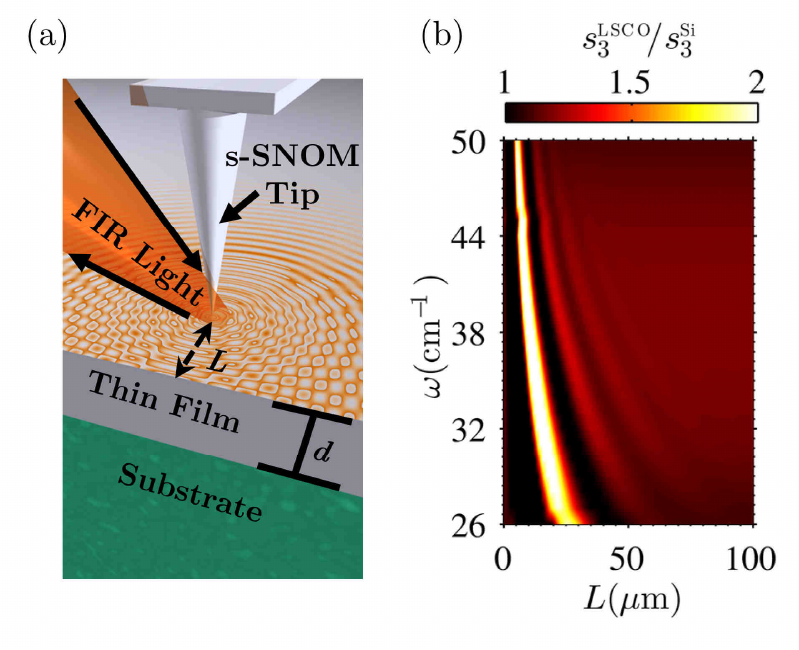}
\caption{(a) Diagram of a proposed SPI experiment. (b) Simulated $s_3$ signal on an LSCO thin film ($x=0.15$) on a Si substrate. The periodic fringes are formed by constructive interference of superfluid plasmons after reflection off the sample edge at $L=0$. Simulation parameters: $d$=10\,nm,  $T=5$K, tip radius $\approx1\,\mu m$.
\label{fig:schem}}
\end{figure}

 Figure~\ref{fig:schem}(a) shows the schematics of a proposed \mbox{SPI} experiment. The sample is a superconducting film of thickness $d$ on a suitable substrate. Infrared radiation is focused onto the metallic tip of an s-SNOM and the tip is raster scanned over the sample near a physical or electronic boundary. High-$q$ collective superfluid excitations launched by the tip propagate to and reflect off the boundary, then travel back towards the tip to form a standing wave. The maxima, or fringes, of this standing wave are separated by a distance $\lambda_m/2$, where $\lambda_m$ is the mode wavelength. By scanning the tip perpendicular to the boundary, one can image these fringes in real space. These images can then be used to extract the mode wavevector as described below.

Figure~\ref{fig:schem}(b) shows a simulation of the interference fringes on an LSCO thin film for varying excitation frequencies. In this figure, $L$ is the position of the tip relative to the boundary in the film, and the vertical axis is the frequency of incident light. We use the measured optical constants of LSCO ($x=0.15$, $T\ll T_c$) to compute the tip-scattered electric field. We model the tip as a metallic spheroid.~\cite{Fei2012a,Zhang2012} The detected s-SNOM signal is $s^{\mathrm{LSCO}}_3$, the amplitude of the field scattered from an s-SNOM tip over LSCO demodulated at the third harmonic of the tip tapping frequency. We report the detected signal normalized to $s^{\mathrm{Si}}_3$, the spectrally flat reference signal from bulk undoped silicon. In the calculations presented here, we assume Si to be the substrate, but any appropriate substrate could take its place. We also use a spheroid with apex dimensions chosen for maximum coupling efficiency to the surface modes at these frequencies, such that the tip radius $a=1/q_m\approx 1 \mu\textrm{m}$. This radius is larger than a typical s-SNOM tip but still allows for sub-diffractional spatial resolution at THz frequencies.

The fringe pattern in Fig.~\ref{fig:schem}(b) can be used to extract the tip-excited superfluid mode wavevector $q_m$. Following Ref.~\citen{Fei2012a}, we write 
\begin{equation}
q_m=\dfrac{2\pi}{\lambda_m}\left(1+i\gamma_m\right)\, 
\end{equation}
where $\gamma_m=\textrm{Im }(q)/\textrm{Re }(q)$ is a dimensionless damping parameter that determines the propagation length of the mode in real space. In Fig.~\ref{fig:schem}(b), the fringe periodicity is $\lambda_m/2$, and the decay in fringe amplitude is governed by $\gamma_m$. Thus the near-field image can be used to evaluate both the real and the imaginary parts of $q_m$ at a given excitation frequency, provided enough fringes are visible to accurately extract both $\lambda_m$ and $\gamma_m$. By varying the excitation frequency, or by illuminating with a broad-band source, one can map the entire dispersion of collective modes in the film. For reasons discussed in Sec.~\ref{sec:math}, the symmetric SPP and the principal HWM will be the most prominent.

Spatial variations in $q_m$ are related to inhomogeneities in the optical properties of the film, similar to those seen in near-field images of graphene~\cite{Fei2013}. The shortest length scale over which SPI can determine $q_m$ is approximately equal to $\lambda_m$. As is evident from Fig. \ref{fig:lscorp}(a), $\lambda_m$ is much smaller than the wavelength of light in vacuum, allowing SPI to conduct subdiffraction-limited measurements. For a 10-nm film of LSCO deposited on silicon $\left(\epsilon_2=11.7\right)$, we find the \mbox{SPI} spatial resolution to be approximately $\lambda_0/10$, where $\lambda_0$ is the free-space wavelength. One could achieve even higher spatial resolution by increasing $\epsilon_2$, which decreases $\lambda_m$. A further advantage of SPI is that it generates tip-launched surface modes, as opposed to edge-launched modes in previous s-SNOM based surface polariton measurements.~\cite{Huber2005} This means that SPI can investigate surface mode dispersions without the need for additional structure fabrication on the sample. Moreover, tip-launching provides SPI with the ability to directly image both physical and electronic boundaries in the sample.~\cite{Fei2013}

\begin{figure}
\includegraphics{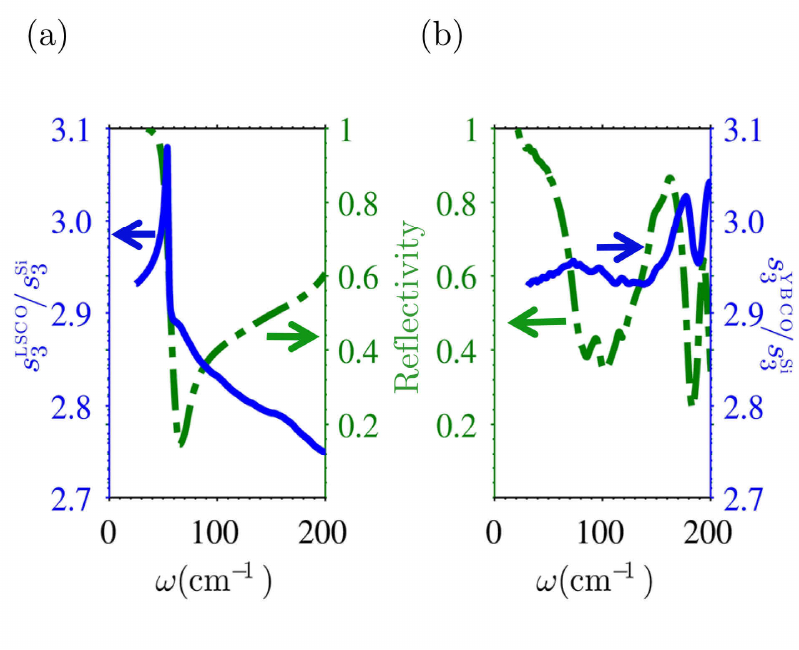}
\caption{Calculated $s_3$ (solid blue outer axis) and the magnitude of the far-field $c$-axis reflectivity (dashed green inner axis) of (a) an LSCO ($x=0.15$, $T\ll T_c$) crystal with a thickness of 1 mm using phonon-subtracted optical constants,~\cite{Tajima2005,Uchida1996} and (b) a YBCO ($x=6.75$, $T\ll T_c$) crystal with a thickness of 1 mm using measured optical constants.~\cite{Lee2005, Homes1995} The far-field reflectivity data for LSCO are reproduced from Ref.~[\citen{Uchida1996}]; data for YBCO, from Ref.~[\citen{LaForge2007}]. The tip radius is 10 nm.
\label{fig:htcspec}}
\end{figure}

\section{\lowercase{s}-SNOM Spectroscopy of Superconductors}
\label{sec:spec}

s-SNOM methods~\cite{Pohl1984, Denk1991, Cvitkovic2007, Amarie2009,Novotny2006, Atkin2012} provide information about the electromagnetic response of the sample on length scales equal to the tip radius,~\cite{Taubner2003, Keilmann2004,Huber2005,Moon2012b} typically $10$--$30\,\mathrm{nm}$. The ultrahigh spatial resolution is also an advantage of spectroscopic s-SNOM measurements.~\cite{Taubner2004,Brehm2006} In Fig.~\ref{fig:htcspec}, we apply the same spheroid model as above~\cite{Zhang2012} to calculate the spectrum of the s-SNOM amplitude $s_3$ from both an LSCO ($x=0.15$) and a YBCO ($x=6.75$) crystal with the tip above the exposed surface of the $ab$-plane. Unlike the calculations shown in Fig.~\ref{fig:schem}, the $s_3$ spectra in Fig.~\ref{fig:htcspec} are calculated for the tip far from any boundaries in the sample. In the LSCO crystal [Fig.~\ref{fig:htcspec}(a)], there is a sharp peak in the $s_3$ spectrum at \mbox{~55 cm\textsuperscript{-1}}, the same frequency as the asymptotic SPP mode in Fig.~\ref{fig:lscorp}. As described above, this surface mode is due to the Josephson plasma resonance of superfluid current along the $c$-axis of the sample. It is instructive to compare the $s_3$ resonance at $\omega_c$ with the far-field $c$-axis reflectivity of the same sample, which we reproduce here from Ref.~\citen{Uchida1996}. The $s_3$ spectrum resonance peak at $\omega_c$ coincides with the Josephson plasma edge in reflectivity.

\begin{figure}
\includegraphics{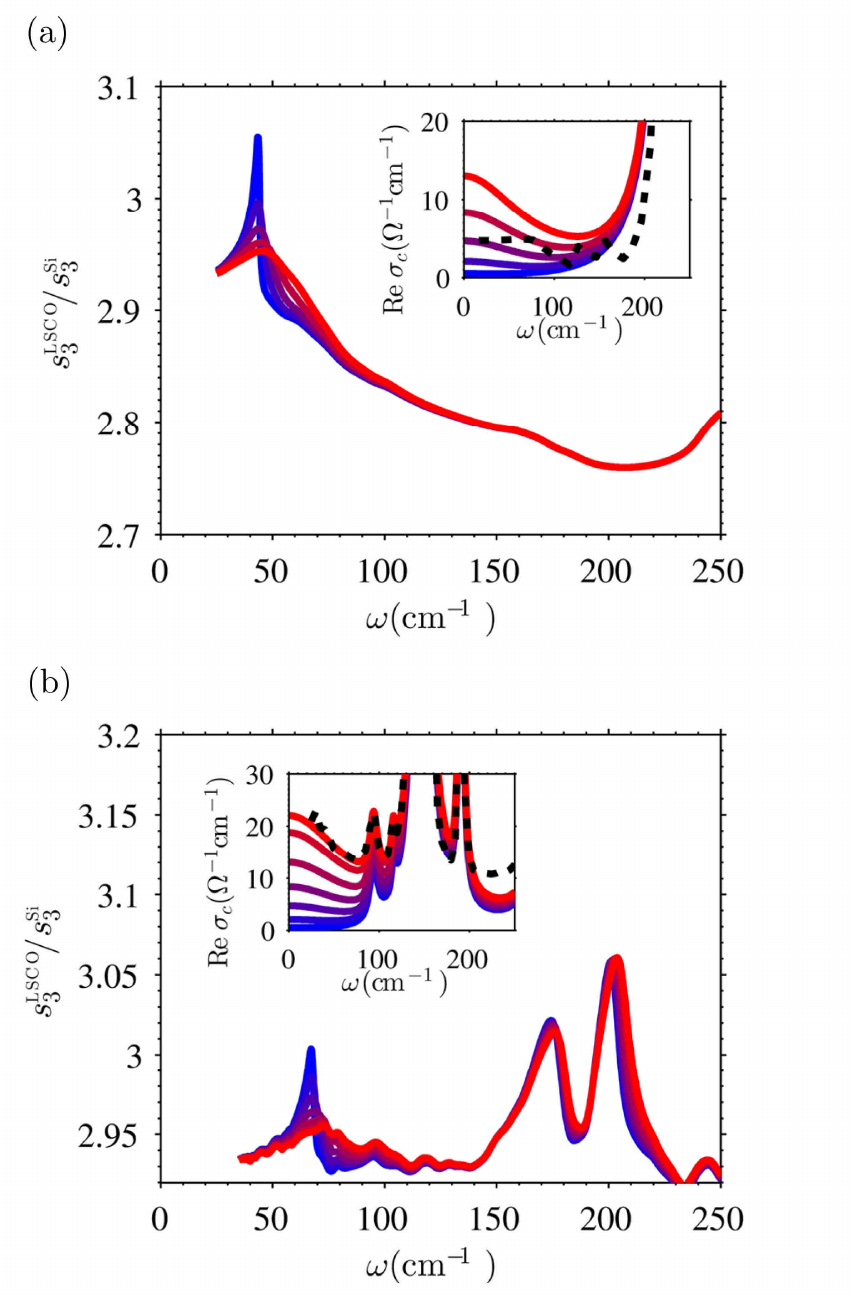}
\caption{(Calculated $s_3$ for anisotropic superconducting (a) LSCO ($x=0.15$, $T\ll T_c$) and (b) YBCO ($x=6.75$, $T\ll T_c$) thin films of thickness $1\,\mu\mathrm{m}$. (a) The LSCO in-plane optical constants are taken from Ref.~\cite{Tajima2005}. The $c$-axis optical constants are modeled with Eq.~\eqref{eq:twofluid} with parameters $\bar{\omega}_c=205\,\mathrm{cm}^{-1}$, $\beta=80\,\mathrm{cm}^{-1}$, $\epsilon_{\infty}=4$, $\omega_0=240\,\mathrm{cm}^{-1}$, $\omega_p=1000\,\mathrm{cm}^{-1}$, $\gamma=10\,\mathrm{cm}^{-1}$. Different traces are for different values of $\omega_n$ in this model. As $\omega_n$ increases, damping at low frequencies broadens the Josephson plasma resonance feature in $s_3$. Inset: Real part of $\sigma_c$ used in calculating $s_3$ in the main panel. The dashed black line is the real part of $\sigma_c$ for LSCO ($x=0.17$, $T\ll T_c$), taken from reflectivity data.~\cite{Dordevic2003} (b) Same as (a), but for YBCO and with two-fluid model parameters $\bar{\omega}_c=370\,\mathrm{cm}^{-1}$, $\beta=80\,\mathrm{cm}^{-1}$, $\epsilon_{\infty}=14$, and six Lorentzian oscillators at $\omega_0=94,\,116,\,146,\,191,\,280$ and $310\,\mathrm{cm}^{-1}$. In the inset, the dashed black line is the measured $c$-axis conductivity for this sample.~\cite{Homes1995}}
\label{fig:specdamp}
\end{figure}
\begin{figure}
\includegraphics{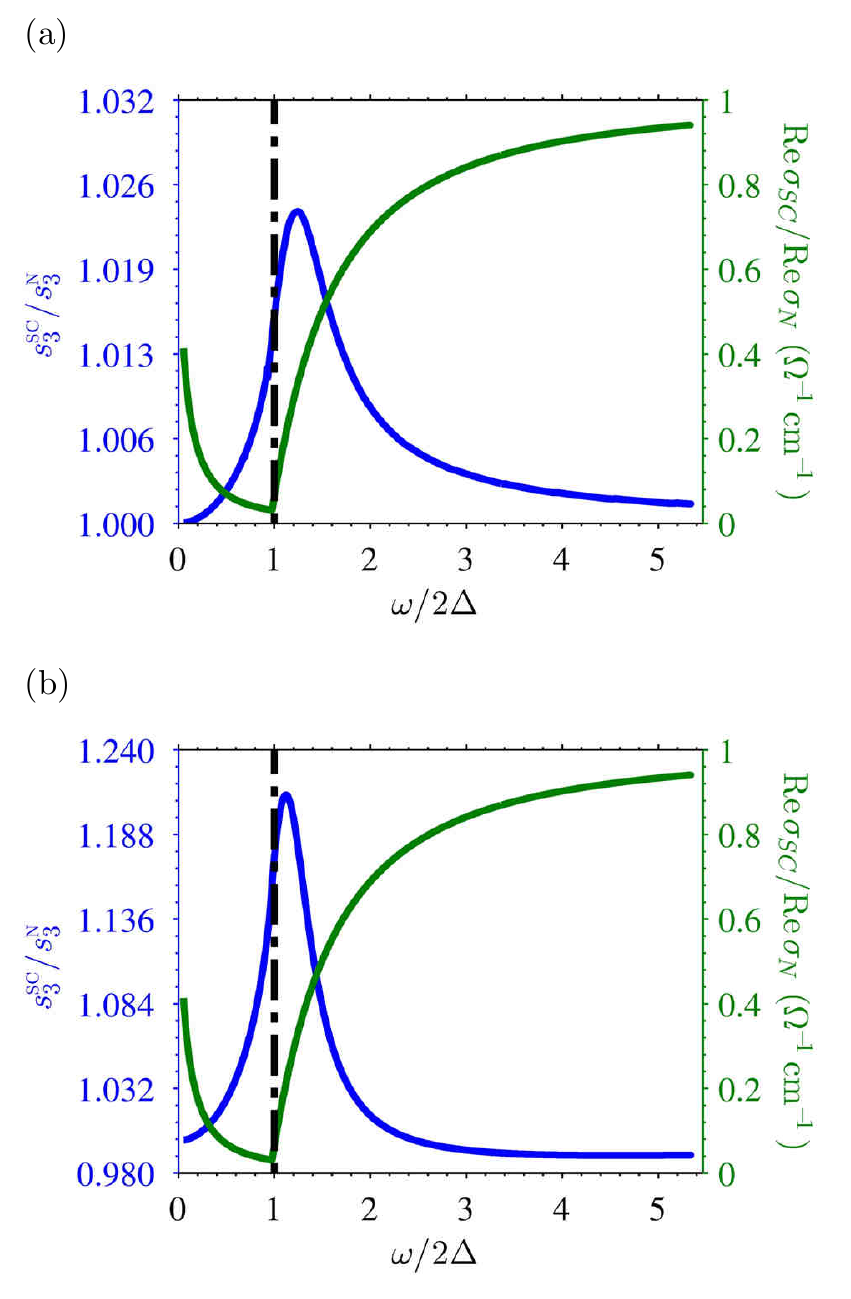}
\caption{Calculated $s_3$ ratio in the superconducting (SC) and normal (N) state of a BCS thin film. (a, b) There is a sharp peak in the $s_3$ spectrum (blue line, left axis) slightly above the gap due to the increased surface plasmon dissipation above $2\Delta$ (see text). The solid green line (right axis) is the ratio of the real parts of $\sigma$ in the SC and N states. In (a), the optical constants used are those of a good metal in the dirty limit ($\omega_p=62,000\,\mathrm{cm}^{-1}, 1/\tau=1450\,\mathrm{cm}^{-1}$) and a tip radius of 500 nm, while (b) uses a much lower plasma frequency but a similar scattering rate ($\omega_p=8,200\,\mathrm{cm}^{-1}, 1/\tau=2000\,\mathrm{cm}^{-1}$) and a tip radius of 50 nm. In (a) and (b), $d=5\,\mathrm{nm}$, $T_{SC}=0.5T_c$ and $T_N \approx T_c$.}
\label{fig:bcsspec}
\end{figure}

Figure~\ref{fig:htcspec}(b) shows the calculated $s_3$ spectrum and the measured far-field $c$-axis reflectivity for an underdoped YBCO ($x=6.75$) crystal. In YBCO, the peak in $s_3$ at the Josephson plasma frequency is much broader and of lower amplitude. Underdoped YBCO has a much higher residual $c$-axis conductivity at low frequencies than optimally doped LSCO [Figs.~\ref{fig:lscorp}(a) and\ref{fig:ybcorp}(a)]. While the s-SNOM response is strictly a function of $r_P$, which depends on both in-plane and $c$-axis optical constants, our modeling suggests that the Josephson resonance peak in $s_3$ depends strongly on Re~$\sigma_c\left(\omega\approx\omega_c\right)$, the real part of the $c$-axis conductivity in the superconducting state at the Josephson plasma frequency. To demonstrate this fact, we calculated $s_3$ spectra for LSCO samples with differing $c$-axis residual conductivities. We use a two-fluid model augmented by several Lorentzians (representing phonon resonances) to capture the essential features of the low-frequency $c$-axis optical constants:
\begin{equation}
\epsilon_c\left(\omega\right)=\epsilon_{\infty}-\frac{\bar{\omega}_c^2}{\omega^2}-\frac{\omega_n^2}{\omega^2+i\omega\beta}+\sum_n{\frac{\omega_{pn}^2}{\omega_{0n}^2-\omega^2-i\omega\gamma_n}}
\label{eq:twofluid}
\end{equation}
where $\bar{\omega}_c$ is the unscreened Josephson plasma frequency; $\omega_n$ and $\beta$ are the normal fluid plasma and scattering frequency, respectively; and $\omega_{pn}$, $\omega_{0n}$, and $\gamma_n$ are the $n$th Lorentzian's oscillator strength, center frequency, and broadening, respectively. We vary the residual conductivity by varying $\omega_n$, as shown in the insets in Fig.~\ref{fig:specdamp}. 

Figure~\ref{fig:specdamp}(a) shows the calculated $s_3$ spectrum for an LSCO sample with $ab$-plane optical constants taken from reflectivity measurements ($x=0.15$, $T\ll T_c$),~\cite{Tajima2005} and varying $c$-axis optical constants calculated using Eq.~\eqref{eq:twofluid}. As the residual $c$-axis conductivity increases, the Josephson feature in the $s_3$ spectrum broadens. The same behavior is shown in Fig.~\ref{fig:specdamp}(b) for YBCO ($x=6.75$, $T\ll T_c$). While both these samples have different in-plane optical constants, we find that in both cases the Josephson plasma resonance in the $s_3$ spectrum is no longer distinguishable from the background when the residual $c$-axis conductivity exceeds $\sim 5$ to $8~\Omega^{-1}\textrm{cm}^{-1}$. In materials satisfying this rough constraint, s-SNOM spectroscopy can directly measure the Josephson plasma frequency $\omega_c$. By raster scanning the s-SNOM tip and measuring a spectrum at every point,~\cite{Bouillard2010} one could measure spatial variations in the $c$-axis superfluid density at nanometer length scales. This spatial resolution is at least an order of magnitude higher than what is currently achievable with other local measurements of superfluid density.~\cite{Moler1998, Luan2011, Lee2003}

We have also calculated the $s_3$ spectrum for a typical BCS-like superconductor. In Fig.~\ref{fig:bcsspec}(a), we show the $s_3$ spectrum for a superconducting 10-monolayer Pb thin film, normalized to $s_3$ of the same film in the normal state. We model the temperature-dependent dielectric function of the thin film with a BCS model~\cite{Zimmermann1991} using Drude parameters measured by infrared spectroscopy~\cite{Pucci2006} and superconducting gap values measured by scanning tunneling microscopy.~\cite{Eom2006} The $s_3$ spectra in the superconducting state exhibits a sharp peak at a frequency slightly above the gap. We observe that the magnitude of this peak increases with decreasing normal state plasma frequency [Fig.~\ref{fig:bcsspec}(b)]. We also observe (not shown) that the magnitude of the $s_3$ peak first increases and eventually shifts to lower frequencies with increasing tip radius. All of these above facts lead us to the following plasmonic interpretation of the origin of the spectroscopic s-SNOM peak near $2\Delta$. 

As discussed above, features in $s_3$ spectra are related to the reflection coefficient $r_P$ of the sample. BCS thin films have a surface plasmon mode\cite{Buisson1994} that is practically lossless at frequencies $\omega<2\Delta$, and becomes broadly dissapative at frequencies $\omega>2\Delta$ [cf. Fig~\ref{fig:bcsrp}(c)]. The $s_3$ peak at $\omega\approx2\Delta$ is then a result of the rapid ``switching-on" of the broadened plasmon damping above the gap.\cite{BCSdispnote} The superconducting gap in $s_3$ spectra should be more visible in superconductors in the dirty limit with low normal-state plasma frequencies, where the difference between superconducting- and normal-state dissipation is highest.

We therefore conclude that s-SNOM spectroscopy at low frequencies will allow for nanoscale spatial resolution of the superconducting energy gap in thin-film samples. While scanning tunneling microscopy allows for atomic-scale spatial resolution of the energy gap, acheivable scan ranges are less than a micron. Current low-temperature s-SNOMs can image areas of $2,500\,\mu\mathrm{m}^2$. s-SNOM will allow for large-scale imaging of the superconducting gap with a high spatial resolution, adding to the mesoscopic picture of the phase transition and enhancing our understanding of phase separation in unconventional superconductors.

\section{Conclusion and Outlook}
\label{sec:conc}

We have shown that near-field imaging and spectroscopy are viable methods for probing properties of the superconducting condensate in both conventional and layered superconductors with a high spatial resolution. In layered superconductors, we predict that s-SNOM SPI methods can image surface superfluid modes in thin films or exfoliated crystals, and that s-SNOM spectroscopy can measure the $c$-axis superfluid density in macroscopic samples. In conventional superconductors, we predict that s-SNOM spectroscopy can accurately determine the magnitude of the superconducting gap. All of these measurements can be done with sub-diffractional to nanoscale spatial resolution. Recent demonstrations of both low-temperature~\cite{Yang2013} and THz~\cite{Moon2012b} s-SNOM attest to the feasibility of the experiments we propose above. Our proposal could shed new light on the nature of spatial inhomogeneities in anisotropic superconductors. Furthermore, in the frequency range studied above cuprate superconductors are ``natural'' realizations of so-called hyperbolic materials.~\cite{Guo2012,Poddubny2013} The guided collective modes present in such materials hold promise for device applications in super-resolution imaging, sensing, and nanolithography.~\cite{Guo2012,Dai2013} Additionally, the techniques we have described could possibly be extended to observe exotic collective modes due to the broken symmetry in the superconducting state.~\cite{Goldman2006,Podolsky2011,Endres2012,Orenstein2007}
\\

Work by H.T.S., Z.F., B.C.C., A.S.M., and D.N.B. was supported by Grant. No. NSF-1005493. Work by J.S.W., B.Y.J., and M.M.F. was supported by ONR and UCOP. Work by Y.S.L. was supported by Grant No. NSF-2013R1A2A2A0106856.

\bibliography{Stinson2014}

\begin{thebibliography}{59}
\bibitem [{asy()}]{asympnote}%
  \BibitemOpen
  \href@noop {} {}\bibinfo {note} {For many underdoped cuprates, and
  specifically for the materials we consider in this work,
  $\gamma\sim100$-$1000$. For that reason, we use $\omega_{as}$ and $\omega_c$
  interchangeably.}\BibitemShut {Stop}%
\bibitem [{BCS()}]{BCSdispnote}%
  \BibitemOpen
  \href@noop {} {}\bibinfo {note} {As long as $q_{tip}\gg q_{cross}$ where
  $q_{cross}=q_{sp}\left(\omega=\omega_g\right)$ is the plasmon wavevector at
  the gap frequency, the s-SNOM gap peak will not depend on the film thickness.
  However, as the film thickness increases $q_{cross}$ decreases, resulting in
  a decreased magnitude of the s-SNOM peak.}\BibitemShut {Stop}%
\end{thebibliography}

\end{document}